\documentclass[aps,pra,showpacs,twocolumn,eqsecnum,superscriptaddress]{revtex4}
\usepackage{graphicx}
\usepackage{amsmath,amssymb,amsthm,dsfont,bm}

\begin{document}
\preprint{}
\title{Mapping of the 2+1 q-deformed Dirac oscillator onto the q-deformed Jaynes-Cummings model: It's non-relativistic limit and {\it Zitterbewegung} effect}
\author{Parisa Majari} 
\email{maajari@gmail.com}
\affiliation{Department of Science, University of Kurdistan, Sanandaj, Iran}
\author{Alfredo Luis}
\email{alluis@fis.ucm.es}
\affiliation{Department of Optics, University Complutense, Madrid, Spain}
\author{Mohammad Reza Setare}
\email{rezakord@ipm.ir}
\affiliation{Department of Science, University of Kurdistan, Sanandaj, Iran}

\date{\today}

\begin{abstract}
We develop the equivalence between the two-dimensional Dirac oscillator and the anti-Jaynes-Cummings model within a q-deformed 
scenario. We solve the Hamiltonian spectrum and the time evolution for number and coherent initial states, focusing on the appearance 
of the {\it Zitterbewegung} effect. We show the lack of preservation of the q-deformed versions of the total angular momentum that reproduces 
a collapse-revival structure. We provide suitable relations for the non relativistic limit. 
\end{abstract}


\maketitle

\section{Introduction}
The introduction and proper combination of suitable physical models provide a significant advance in many different areas of physics. 
In this work we combine the relativistic Dirac oscillator, the Jaynes-Cummings model, and the idea of q-deformation. 

The Dirac oscillator provides a generalization to the relativistic domain of one of the most applied models in classical and quantum physics
\cite{IMC67,MS89}. This is a notably simple solvable equation, linear both in momentum and position, formally looking as a simple minimal 
coupling of a charged particle with a magnetic field.  Accordingly, it has found applications in nuclear physics, quantum chromodynamics, 
many-body theories, supersymmetric relativistic quantum mechanics and quantum optics \cite{BMS07,BML08,SL10,FSBKMS13}.

This relativistic generalization is accompanied of new effects such as the unavoidable emergences of spin, negative energies and rather 
unexpected behaviors such as the Klein tunneling and the {\it Zitterbewegung} as the result of interference between positive- and 
negative-energy components. A further key feature, is that the Dirac oscillator contains in itself another germane model in quantum physics: 
the Jaynes-Cummings model (JC), or, more rigorously, the anti-Jaynes-Cummings version (AJC)  \cite{LCLJ11,JMD12}. This describes 
the coupling of a discrete spin-like variable with a bosonic degree of freedom \cite{JCM}. The powerfulness of the model relies on its 
simplicity and its capacity to embrace very different physical situations such as light-matter interaction, dynamics of trapped ions, or Bose-Einstein 
condensates \cite{LBMW03}. In particular this model allows to use quantum-optics based technology or  trapped ions technology for 
simulating relativistic quantum mechanics  \cite{BMS07,BML08}.

On the other hand, q-deformation provides a quite simple but rich and powerful enough way to describe complex situations such as electronic 
conductance in disordered metals and doped semiconductors, phonon spectrum in ${}^4$He,  oscillatoryÐrotational spectra of diatomic and 
multi-atomic molecules \cite{qD,EM06,LRW97}. Deformation is accomplished by entering a dimensionless parameter $q$ in normal Weyl 
algebra. This grants a more general case of the ancient theory at the same time it replicates the primary theory for $q=1$. For example,  we may 
say that relativistic quantum mechanics is a q-deformed Newtonian quantum mechanics, where the deformation parameter is the velocity ratio
$q=v/c$. Whenever $q$ tends to zero, relativistic quantum mechanics convert to the Newtonian quantum mechanics.

Thus, in this work we investigate the q-deformed perspective of the interplay between Dirac oscillator and Jaynes-Cummings model following 
Ref. \cite{HS16}, both in the relativistic domain and in the non-relativistic limit.  In particular we focus on the preservation in the q-deformed 
scenario of peculiar relativistic features such as the {\it Zitterbewegung} effect. 

\section{The $\mathrm{q}$-deformed model}

Let us present the main results regarding the q-deformation and the equivalence proposed.

\subsection{$\mathrm{q}$-Deformed observables and states}

A general case of  modified Weyl algebra holds as follows \cite{LRW97,VD02}:
\begin{equation}
\label{cr}
{a_{q}}a_{q}^\dag-qa_{q}^\dag a_{q}=1 , \quad \mathrm{for} \;\; \; q \leq 1, 
\end{equation}
where $a_{q}$ and ${a_{q}}^\dag$ are annihilation and creation operators, that operate such that:
\begin{equation}
{a_{q}}|n \rangle =\sqrt{[n]}|n-1\rangle ,  \quad {a_{q}^\dag}|n\rangle =\sqrt{[n+1]}|n+1\rangle ,
\end{equation}
where the number basis $|n \rangle $ is actually defined by the eigenvalue equation 
\begin{equation}
\label{[num]}
[ \hat{n} ] |n \rangle = [n] |n \rangle  , \quad [ \hat{n} ]  \equiv a^\dagger_q a_q ,
\end{equation}
and $[n]$ satisfies this relation \cite{EM06}:
\begin{equation}
[n]=\frac{1-q^n}{1-q} .
\end{equation}
For $q \rightarrow 1$ naturally $[n] \rightarrow n$ while for $q  < 1$ we have $[n]  <  n$, with  $[n] \rightarrow 1$ as $q \rightarrow 0$.

Alternatively to the q-number operator $[ \hat{n} ] $ in Eq. (\ref{[num]}) we can define also an operator $\hat{n}$ 
satisfying the standard commutation relation with $a_q$, i. e., 
\begin{equation}
\label{num}
\hat{n}  |n \rangle = n |n \rangle  , \quad [a_q, \hat{n} ] =  a_q .
\end{equation}

We can present two alternative realizations of the q-deformed algebra. On the one hand we have the following differential 
form for $a_q$ 
\begin{equation}
\label{df}
a_q={ e^{-2i\alpha z}-e^{i \alpha {d \over dz }} {e^{-i \alpha z}}\over {-i\sqrt{1-e^{-2 \alpha^2}}}}  ,
\end{equation}
where $\alpha = \sqrt{- \ln q/2}$ and $z$ is a Cartesian coordinate. On the other hand, after \cite{VD02} we may express 
$a_q$ also as 
\begin{equation}
a_q = F \left (a^\dagger a \right ) a, \quad  F \left ( n \right ) = \sqrt{\frac{[n+1]}{n+1}}  ,
\end{equation}
where naturally $a= a_{q=1}$. In particular this means that q-deformation implies a highly nonlinear interaction. The degree of 
nonlinearity may be reduced to more simple levels if $q \rightarrow 1$. Thus, considering $q = \exp (- \epsilon )$, a 
power series expansion as $\epsilon \rightarrow 0$ leads to
\begin{equation}
\label{li}
F \left (a^\dagger a \right ) \simeq 1 - \frac{\epsilon}{4} a^\dagger a .
\end{equation}
Therefore, from a quantum-optical perspective q-deformation  corresponds to optics in nonlinear media \cite{CR11}.
Also, the limit (\ref{li})  recalls the dynamics of trapped ions where $a$ describes the motion of the center of mass, as 
shown for example in Ref. \cite{VMF95}.

On the other hand, in the opposite limit of large nonlinearity $q \rightarrow 0$ we get
\begin{equation}
F \left (a^\dagger a \right ) \rightarrow \frac{1}{\sqrt{1+a^\dagger a}} , \quad a_q \rightarrow  \frac{1}{\sqrt{1+a^\dagger a}} a ,
\end{equation}
so that
\begin{equation}
a_q | n \rangle \rightarrow   | n-1 \rangle , \quad a_q | 0 \rangle=   0 .
\end{equation}
This is to say that in this limit  $a_q$ becomes the Susskind-Glogower phase operator \cite{RL95}. That is that $a_q$ tends to 
represent the complex exponential of the oscillator phase.

Besides the number states (\ref{[num]}) we consider some analogs of the coherent states defined via the eigenvalue equation \cite{FI91}:
\begin{equation}\label{36}
a_q |\alpha \rangle =\alpha |\alpha \rangle ,
\end{equation}
being 
\begin{equation}\label{cs}
|\alpha \rangle= \frac{1}{\sqrt{e_q (|\alpha |^2)}} \sum_{n=0}^\infty  \frac{\alpha^n}{[n]!} | n \rangle, 
\quad 
|\alpha|^2 < \frac{1}{1-q} ,
\end{equation}
where $[n]! = [n][n-1] \ldots [1]$ with $[0]! = 1$ and the following relation has been used
\begin{equation}\label{38}
{e_{q}(z)}=\sum_{n=0}^{\infty} {z^n \over {[n]!}} .
\end{equation}
These relations hold provided that $|\alpha|^2 < 1/(1-q)$ as a necessary condition such that the number coefficients in Eq. (\ref{cs}) tend to zero 
as $n$ tend to infinity.

Regarding the number statistics let us show that these state are sub-Poissonian for the q-deformed number operator $[ \hat{n} ] $ since 
\begin{equation}
\langle \alpha | [ \hat{n} ] | \alpha \rangle = | \alpha |^2 , \quad 
\langle \alpha | [ \hat{n} ]^2 | \alpha \rangle = | \alpha |^2 + q  | \alpha |^4 .
\end{equation}
We can introduce the Mandel Q parameter in the q-deformed scenario as
\begin{equation}
Q_q = \frac{ \Delta^2 [ \hat{n}] }{\langle [ \hat{n}] \rangle} - 1 = (q-1) | \alpha |^2 ,
\end{equation}
so that $Q_q < 1 \leftrightarrow q <1$. On the other hand, regarding the other number operator  $\hat{n} $ in Eq. (\ref{num}) we have obtained 
numerically that the q-deformed coherent states (\ref{cs}) are always super-Poissonian, that is  
\begin{equation}
Q = \frac{ \Delta^2  \hat{n}  }{\langle \hat{n} \rangle} - 1 > 0 .
\end{equation}
This is illustrated in Fig. 1 where we plot both Mandel parameters $Q_q$, $Q$ as functions of $| \alpha |^2$ for $q=0.25$ (solid line) and  $q=0.75$ (dashed line).

\begin{figure}
\begin{center}
\includegraphics[width=8cm]{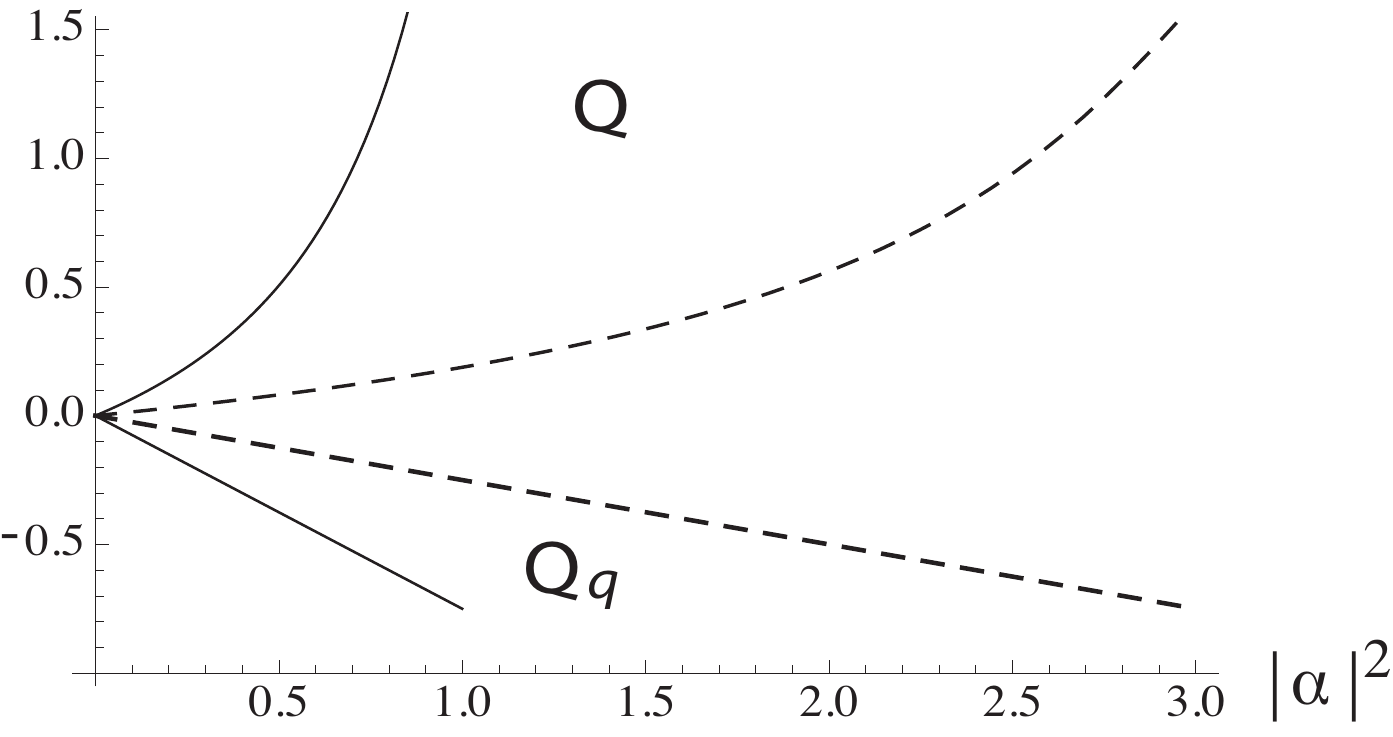}
\end{center}
\caption{Mandel parameters $Q_q$, $Q$ as functions of $| \alpha |^2$ for $q=0.25$ (solid line) and  $q=0.75$ (dashed line). }
\end{figure} 

\subsection{The $\mathrm{q}$-deformed Dirac oscillator}

The Hamiltonian of the Dirac oscillator in two spatial coordinates $x, y$ is given by  \cite{HS16}
\begin{equation}\label{HD}
{\
H^D}=\begin{bmatrix}
mc^2 &i\sqrt{4m c^2 \hbar \omega} \, a_\ell^\dag  \\
-i \sqrt{4m c^2 \hbar \omega} \, a_\ell & -mc^2
\end{bmatrix},
\end{equation}
where $m$ is the rest mass of the electron, $\omega$ is the Dirac-oscillator frequency,  $c$ is the speed of light, and 
$a_\ell$ is the left-handed combination of $x,y$ dynamics \cite{BMS07}
\begin{equation}\label{5}
a_{\ell}={a_{x}+ia_{y}\over{\sqrt{2}}} .
\end{equation}

Now we can consider  the q-deformed version of the above Hamiltonian \cite{HS16}:
\begin{equation}
{\
H^D_{q}}=\begin{bmatrix}
mc^2 &i \sqrt{4m c^2 \hbar \omega} \, a_{\ell,q}^\dag  \\
- i \sqrt{4m c^2 \hbar \omega} \, a_{\ell , q} & -mc^2
\end{bmatrix},
\end{equation}
or equivalently,
\begin{equation}\label{HDq}
H^D_{q} = mc^2 \sigma_z + i \sqrt{4m c^2 \hbar \omega} \left ( \sigma_+ a^\dagger_{\ell , q}  -   \sigma_- a_{\ell , q}  \right ) ,
\end{equation}
where $\sigma$ refers to the corresponding Pauli matrices.  Alternatively, 
\begin{equation} \label{HDq2}
{\
H^D_q}=mc^2\sigma_{z}+i \sqrt{4m c^2 \hbar \omega}\begin{bmatrix}
0 & a_{x, q}^\dag-ia_{y, q}^\dag  \\
-a_{x, q}-ia_{y, q} & 0
\end{bmatrix} .
\end{equation}

Within the q-deformed scenario we can define the $z$ components of the orbital $\bm{L}$, spin $\bm{S}$, and total $\bm{J}$ 
angular momentum operators:
\begin{equation}
\label{amo}
L_z = \hbar  \left ( a^\dagger_{r,q} a_{r,q} - a^\dagger_{\ell,q} a_{\ell,q} \right ) ,\quad S_z = \frac{\hbar}{2} \sigma_z , \quad J_z = L_z + S_z ,
\end{equation}
where  the right-handed mode $a_{r,q}$ is defined as 
\begin{equation}\label{5}
a_{r,q}={a_{x,q} - ia_{y, q} \over{\sqrt{2}}}  ,
\end{equation}
and we will always assume that this mode is in the vacuum state. 

Next we compare the q-deformed  Dirac oscillator $H_q^D$ with  the q-deformed Jaynes-Cummings and 
anti-Jaynes-Cummings models. 

\subsection{The q-deformed JC and AJC Hamiltonians}

The JC model is an example of Rabi model in the framework of  quantum electrodynamics that presents the coupling of two 
energy levels, described by the Pauli matrices $\sigma$, with an harmonic oscillator, described by the complex-amplitude 
or annihilation operator $a$. As two typical realizations the oscillator we can mention a quantum electrodynamic field mode, i. e., 
photons, or the motion of the center of mass of a trapped ion, that is phonons. This last case is specially interesting since it 
allows to simulate both JC and AJC models simply by properly selecting the frequency tuning of a laser field illuminating the 
ion, leading to the  AJC and JC Hamiltonians \cite{LBMW03}
\begin{equation}
H^{AJC} =  \hbar \delta \sigma_{z}+ \hbar \eta(\sigma^{+} a^{\dag} e^{i\phi}+\sigma^{-} a e^{-i\phi} ) ,
\end{equation}
\begin{equation}
H^{JC}  =   \hbar \delta^\prime \sigma_{z} + \hbar \eta^\prime (\sigma^{+} a e^{i\phi^\prime}+\sigma^{-} a^{\dag} e^{-i\phi^\prime}) ,
\end{equation}
where $\eta, \eta^\prime$ are coupling constant, $\delta, \delta^\prime$ represent detuning, and $\phi, \phi^\prime$ are arbitrary phases. 
  
Now we q-deforme these JC and AJC Hamiltonians simply by substituting $a$, $a^\dagger$ by their the q-deformed versions of creation 
and annihilation operators:
\begin{equation}\label{HAJCq}
H_q^{AJC} =  \hbar \delta \sigma_{z} + \hbar \eta(\sigma^{+} a_q^{\dag} e^{i\phi}+\sigma^{-} a_q e^{-i\phi} )  ,
\end{equation}
\begin{equation} \label{HJCq}
H_q^{JC} =  \hbar \delta^\prime \sigma_{z} + \hbar \eta^\prime (\sigma^{+} a_q e^{i\phi^\prime}+\sigma^{-} a_q^{\dag} e^{-i\phi^\prime}) .
\end{equation}
In general terms, q-deformed JC Hamiltonians have been studied in \cite{CMM94}.

\subsection{Equivalence}

A readily inspection of Eqs. (\ref{HDq}) and (\ref{HAJCq}) reveals an exact mapping between them if we properly 
identify the complex-amplitude operators
\begin{equation}
H_q^D \equiv H^{AJC}_q \quad a_q \equiv a_{\ell,q},
\end{equation}
with the following correspondence of parameters  
\begin{equation}
\hbar \eta =   \sqrt{4m c^2 \hbar \omega}, \quad \phi = \pi/2, \quad \hbar \delta = mc^2 . 
\end{equation}
Alternatively, if we consider explicitly the two-dimensional nature of our Dirac oscillator the equivalence reads
\begin{equation}
H_q^D \equiv H^{AJC}_{q,x} +  H^{AJC}_{q,y} ,
\end{equation}
with 
\begin{eqnarray}
& \hbar \eta_x = \hbar \eta_y =  \sqrt{2 m c^2 \hbar \omega},  & \nonumber \\
&\phi_x = \pi/2,   \quad \phi_y = 0, \quad  \hbar \delta_x = \hbar \delta_y   = mc^2 . &
\end{eqnarray}

\section{Energy spectrum of the $\mathrm{q}$-deformed Dirac oscillator}

Next we derive the energy spectrum of the q-deformed Dirac oscillator $H_q^D$. The spectrum must be calculated from scratch since 
the commutation relations have changed in comparison with the $q=1$ case. The q-deformed time-independent Dirac equation is:
\begin{equation}\label{ti}
H^D_{q}|\psi \rangle =E |\psi \rangle ,
\end{equation}
with $H^D_{q}$ in Eq. (\ref {HDq}). Looking for spinor solutions,
\begin{equation}\label{24}
{\
|\psi  \rangle }=\begin{bmatrix} |\psi_1 \rangle   \\ |\psi_2 \rangle \end{bmatrix},
\end{equation}
we get that Eq. (\ref{ti}) is equivalent to the pair of equations  
\begin{eqnarray} \label{tc}
& m c^2 |\psi_{1} \rangle + i \sqrt{4m c^2 \hbar \omega } a_{\ell,q}^\dag |\psi_2 \rangle = E |\psi_1 \rangle, & \nonumber \\
&  &  \\
& - m c^2 |\psi_{2} \rangle - i \sqrt{4 m c^2 \hbar \omega} a_{\ell,q} |\psi_1 \rangle = E |\psi_2 \rangle. & \nonumber 
\end{eqnarray}
We have a readily solution with $E =E_0= mc^2$,  $|\psi_2 \rangle =0 $,  and $|\psi_1 \rangle = | 0 \rangle$, that is 
\begin{equation}
 |E_0  \rangle = | 0 \rangle   |\chi_{\uparrow}  \rangle .
 \end{equation}
Whenever $E \neq mc^2$ combining Eqs. (\ref{tc}) we get :
\begin{equation}\label{25}
(E^2-{m^2}c^4)|\psi_{1} \rangle ={4m c^2 \hbar \omega} {a_{\ell,q}^\dag} {a_{\ell,q}}|\psi_{1} \rangle ,
\end{equation}
\begin{equation}\label{26}
(E^2-{m^2}c^4)|\psi_{2} \rangle ={4m c^2 \hbar \omega} {a_{\ell,q}} {a_{\ell,q}^\dag} |\psi_{2} \rangle ,
\end{equation}
and we recall that in terms of q-deformed number operators $[\hat{n}] = a_{\ell,q}^\dag a_{\ell,q} $ and  $[\hat{n}+ 1] =a_{\ell,q} a_{\ell,q}^\dag$.
From the above equations the energies can be easily obtained as follows:
\begin{equation}\label{Eq}
E =\pm  E_n=\pm mc^2{\sqrt{1+4\xi[n]}} , \quad \xi= \frac{\hbar \omega }{mc^2},
\end{equation}
for natural numbers $n \neq 0$. The positive and negative of the energy eigenstates $|\pm E_n \rangle $ are:
\begin{equation}\label{eig}
{\
|\pm E_n \rangle }=\begin{bmatrix}
{\sqrt{E_n \pm mc^2 \over 2E_n}}|n  \rangle  \\
\mp i{\sqrt{E_n\mp mc^2 \over 2E_n}}|n  -1  \rangle 
\end{bmatrix} .
\end{equation}
The above result can be rewritten in terms of the eigenstates of $\sigma_z$, $|\chi_{\uparrow}  \rangle =(1, 0)^{\dag}$ 
and $|\chi_{\downarrow} \rangle =(0, 1)^{\dag}$ as:
\begin{equation}
| \pm E_n \rangle =c_\pm  |n \rangle  |\chi_{\uparrow}  \rangle \mp i c_\mp  |n -1 \rangle   |\chi_{\downarrow} \rangle,
\end{equation}
where $c_\pm $ are:
\begin{equation}\label{30}
c_\pm = {\sqrt{E_n \pm mc^2 \over 2E_n}} .
\end{equation}

Let us present an alternative picture of this diagonalization and time evolution as an interference effect \cite{BML08}. 
To this end we express the  total Hilbert space as direct sum of one- and two-dimensional subspaces as
\begin{equation}\label{61}
\textsf{H}={{\oplus}_{n =0 }^\infty}{\textsf{{H}}}_{n}
\end{equation}
where the subspaces $\textsf{H}_{n }$  are defined as:
\begin{equation}\label{span}
\textsf{H}_{n }= \mathrm{span} \{|n  \rangle |\chi_{\uparrow} \rangle ,|n -1 \rangle |\chi_{\downarrow} \rangle \},
\;\;
\textsf{H}_0= \mathrm{span} \{|0  \rangle |\chi_{\uparrow} \rangle  \} .
\end{equation}
Leaving aside the trivial subspace $\textsf{H}_0$, within each of the two-dimensional spaces $\textsf{H}_n$ 
the Dirac oscillator (\ref{HDq}) can be written simply as, in the basis in Eq. (\ref{span}),
\begin{equation}\label{64}
H^D_{n,q} =mc^2(\sigma_z -\sqrt{4 \xi [n ]} \sigma_{y}) .
\end{equation}
This is equivalent to 
\begin{equation}\label{66}
H^D_{n,q}  = E_n e^{-i\theta_n \sigma_{x}} \sigma _z e^{i\theta_n \sigma_{x}}
\end{equation}
where $\tan ( 2 \theta_n) = \sqrt{4 \xi [n ]}$. Thus  the unitary time evolution operator within each subspace $\textsf{H}_{n }$ 
is given by:
\begin{equation}\label{68}
U_{n,q}=e^{-iH^D_{n,q} t/\hbar} = e^{-i\theta_n \sigma_{x}}e^{-i E_n t \sigma _z /\hbar}e^{i\theta_n \sigma_{x}} .
\end{equation}

This allows us to interpret the whole process as a series of  Mach-Zehnder interferometers in parallel, one for each subspace 
$\textsf{H}_{n }$. In this picture the basis vectors $|n  \rangle |\chi_{\uparrow} \rangle$, $|n -1 \rangle |\chi_{\downarrow} \rangle$ 
play the role of input/output beams, the operators $\exp ( \pm i \theta \sigma_x )$ play the role of input/output beam splitters, while 
$\exp (-i E_n t \sigma _z /\hbar )$ represents the phase difference acquired within the two arms of the interferometer. In other words, 
the beam splitters provide the eigenstates of the Hamiltonian, while the phase difference is given by the eigenvalues. This also 
allows suitable simple expressions regarding the approximations involved in the non-relativistic limit to be considered below.

\section{{\it Zitterbewegung} effect}

The {\it Zitterbewegung} effect is a trembling motion of relativistic particles that occurs due to the interference 
between positive and negative energies \cite{GR00}. Here we want to investigate this effect  for the following q-deformed initial 
state:
\begin{equation}\label{isZr}
|\psi_{0} \rangle =|n -1 \rangle   |\chi_{\downarrow} \rangle =i c_- |+E_n \rangle -i c_+ |-E_n \rangle ,
\end{equation}
for $n \geq 1$. The evolution of this state can be obtained by:

\begin{equation}\label{32}
|\psi_{t} \rangle =i c_- {e^{-i\omega_n t}}|+E_n \rangle -i c_+  {e^{i\omega_n t}}|-E_n \rangle ,
\end{equation}
where
\begin{equation}
\omega_n= \frac{E_n}{\hbar} = \frac{1}{\hbar}  mc^2  \sqrt{1+4\xi [n]},
\end{equation}
is the frequency of {\it Zitterbewegung} oscillation. The oscillation can be well appreciated for example in the mean value of the spin 
and angular momentum operators (\ref{amo}):
\begin{equation}\label{ZB1}
{ \langle L_{z} \rangle }_{t}=-\hbar [n -1]-{{4\xi \hbar [n ] }\over {1+4\xi [n ]}}q^{n -1} \sin^{2}(\omega_n t) ,
\end{equation}
\begin{equation}\label{ZB2}
{ \langle S_{z} \rangle }_{t}=- {{\hbar} \over 2}+{{4\xi \hbar [n ] }\over {1+4\xi [n ]}} \sin^{2}(\omega_n t) ,
\end{equation}
and 
\begin{equation}\label{ZB3}
{ \langle J_{z} \rangle }_{t}=- {{\hbar} \over 2}({1+2[n -1]})+{{4\xi \hbar [n ] }\over {1+4\xi [n ]}}{ \sin^{2}}(\omega_nt)(1-q^{n -1}) ,
\end{equation}
or, equivalently, 
\begin{eqnarray}\label{Jz}
& { \langle J_{z} \rangle }_{t}=- {{\hbar} \over 2}({1+2[n -1]})& \nonumber \\
& +{4\xi \hbar  }{ \sin^{2}}(\omega_nt){({{1-q^{n }}){(1-q^{n -1})}} \over {(1-q+4\xi-4\xi {q^{n }})}} .&
\end{eqnarray}
Several interesting features can be noticed in these results: 

i) As an interference effect we can note that the visibility of the time evolution on $\langle L_{z} \rangle$
and $\langle S_{z} \rangle$ depends on $q$ explicitly as well as implicitly thorough the factor $[ n ]$. Since 
$q \leq 1$ and $[n] \leq n$ we get that the visibility for $q \neq 1$ is lesser than in the undeformed case 
$q=1$. So q-deformation diminishes the effect. 

ii) At difference with the undeformed case $q=1$, in the q-deformed scenario  $J_{z}$ is not constant of the motion except for $n=1$, 
as clearly shown in Eqs. (\ref{ZB3}) and (\ref{Jz}). This is because after the commutation relation (\ref{cr}) we have $[H^D_q, J_z ] \neq 0$.

iii) We can recover the constancy of $\langle J_z \rangle$ provided that we consider an alternative expression for $L_z$, this is that 
$L_z \propto - \hbar \hat{n}$, where $\hat{n}$ is defined in Eq. (\ref{num}). Now the commutation relation (\ref{num}) grants that $[H^D_q , 
J_z] = 0$ and $J_z$ is a constant of the motion as in the undeformed case. 

As a further example of  {\it Zitterbewegung}  we continue by considering a coherent initial state (\ref{cs}) so that $|\psi_0 \rangle = 
|\alpha \rangle  |\chi_{\downarrow} \rangle$ leading to
\begin{equation}\label{40}
\langle L_{z} \rangle_t = - \hbar | \alpha |^2 - \frac{\hbar}{e_q (|\alpha |^2)} \sum_{n=0}^\infty  \frac{| \alpha|^{2 n}}{[n]!} q^n S_{n+1}  (t) ,
\end{equation}
\begin{equation}\label{41}
{ \langle S_{z} \rangle }_{t}=- \frac{\hbar}{2} + \frac{\hbar}{e_q (|\alpha |^2)} \sum_{n=0}^\infty  \frac{| \alpha|^{2 n}}{[n]!}S_{n+1} (t)  ,
\end{equation}
\begin{equation}\label{<J_z>}
{ \langle J_{z} \rangle }_{t}=- \frac{\hbar}{2} \left (1+2| \alpha |^2 \right )+ \hbar \frac{1-q}{e_q (|\alpha |^2)} \sum_{n=0}^\infty  \frac{| \alpha|^{2 n}}{[n]!} 
[n] S_{n+1} (t) ,
\end{equation}
where 
\begin{equation}
S_n (t) = \frac{4\xi [n]}{1+4 \xi [n]} \sin^2 (\omega_n  t) .
\end{equation} 

All above quantities satisfy the proper $q=1$ limit, and in particular we have that $ \langle J_{z} \rangle_t $ is constant for $q=1$. On the other 
hand, for $q \neq 1$ these expressions have the typical structure leading to an scenario of collapse and revivals \cite{CR11}. This is clearly 
shown in Fig. 2. 
\begin{figure}
\begin{center}
\includegraphics[width=8cm]{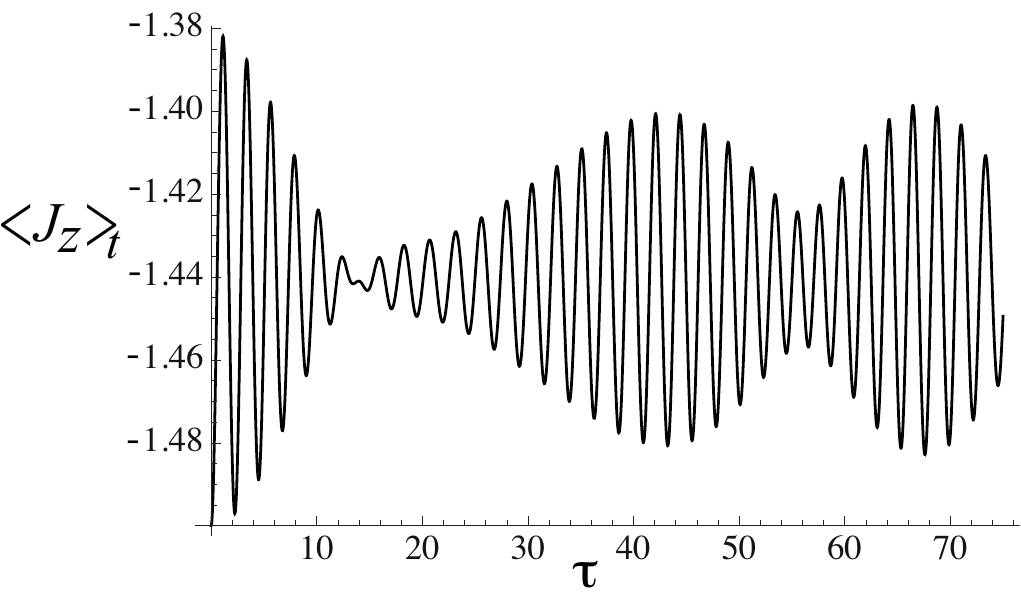}
\end{center}
\caption{Plot of $\langle J_z \rangle_t$ in Eq. (\ref{<J_z>}) as a function of the dimensionless rescaled time $\tau = mc^2 t /\hbar$ for $\alpha = 1$ 
and $q=0.75$. }
\end{figure} 

\section{Non-relativistic limit}

Since typically $\xi [n] \ll m c^2$ it is worth checking the non relativistic limit of the Dirac oscillator, and for that we may use the quasi degenerate 
perturbation theory \cite{CDG04}. In this spirit and given that we know the exact solutions for energies (\ref{Eq}) and states (\ref{eig}), the non 
relativistic limit can be achieved from a series expansion in powers of $\xi$. For the first  order on $\xi$ of the eigenvalues we get:
\begin{equation}\label{50}
E_n \simeq mc^2(1+2\xi [n ]) ,
\end{equation}
while the order $\xi^0$ of the eigenvectors is 
\begin{equation}
\label{aeig}
| + E_n \rangle = |n  \rangle |\chi_{\uparrow} \rangle , \quad | - E_n \rangle = |n -1 \rangle |\chi_{\downarrow} \rangle.
\end{equation}
Thus, within this level of approximation we have $H^D_q \simeq H^D_{\mathrm{eff},q} $ with 
\begin{equation} \label{Heff}
H^D_{\mathrm{eff},q} =\begin{bmatrix}
mc^2+2\hbar\omega a_{\ell, q}^\dag a_{\ell, q} &0  \\
0 & -mc^2-2\hbar\omega a_{\ell, q} a_{\ell, q}^\dag
\end{bmatrix} .
\end{equation}
This result is quite natural since from an optical perspective the condition $\xi [n] \ll m c^2$  actually means a very large detuning. In such a 
case there is no real exchange of photons or phonons and there is just a shift of levels caused by virtual exchanges. Alternatively, from the 
interferometric perspective in this limit  the beam splitters $\exp ( \pm i \theta \sigma_x )$ are replaced by the identity and the phases are 
approximate linearly in $\xi$. 

\subsection{{\it Zitterbewegung} in the non-relativistic limit}

Lets  peruse the existence of {\it Zitterbewegung} in the non-relativistic limit. We can examine it in several steps. On the one hand,
after the form (\ref{Heff}) for the effective Hamiltonian we get that in this non-relativistic limit  
\begin{equation}
[ H^D_{\mathrm{eff},q} , S_z] = 0, \qquad [ H^D_{\mathrm{eff},q} , L_z] = 0 ,
\end{equation}
so that $S_z$ and $L_z$  are constants of the motion and display no  {\it Zitterbewegung}  in this strict limit. 

As a further alternative we may examine the evolution of more sophisticated observables such as say $M = \sigma_- a_{\ell, q}  
+ \sigma_+ a_{\ell, q}^\dagger $. Since the key point of this effect is the interference between negative and positive energies, after 
Eq. (\ref{aeig}) we have that the initial state must be different from (\ref{isZr}). Thus let us consider instead for example the following 
initial state:
\begin{equation}\label{52}
|\psi_{0} \rangle =c_\uparrow |n  \rangle |\chi_{\uparrow} \rangle + c_\downarrow |n -1 \rangle |\chi_{\downarrow} \rangle ,
\end{equation}
where $c_{\uparrow, \downarrow}$ are real constants. It`s time evolution results in:
\begin{equation}\label{53}
|\psi_{t} \rangle = c_\uparrow e^{-i\omega_n t} |n  \rangle |\chi_{\uparrow} \rangle + c_\downarrow e^{i\omega_n t} |n -1 \rangle 
|\chi_{\downarrow} \rangle ,
\end{equation}
where here $\omega_n = mc^2(1+2\xi [n ])/ \hbar $. So we may easily calculate $\langle M \rangle$ leading to 
\begin{equation}\label{54}
\langle M \rangle_t = 2  c_\uparrow  c_\downarrow \sqrt{[n]} \cos \left ( 2 \omega_n t \right ) ,
\end{equation}
displaying the desired oscillating fully analogous  Eqs. (\ref{ZB1}) to (\ref{ZB3}).

On the other hand we may consider also further corrections to the non-relativistic limit. For example, regarding {\it Zitterbewegung} 
we may consider the same initial state (\ref{isZr}) including also the first approximation in powers of $\xi$ in $\exp ( \pm i \theta \sigma_x )$, 
leading  to 
\begin{equation}
\langle J_{z} \rangle_t=- \frac{\hbar}{2} \left ( 1+2[n -1] \right )+4\xi \hbar [n ] (1-q^{n -1})  \sin^2 \left ( \omega_n t \right ) ,
\end{equation}
for the same $\omega_n$ above.

\section{Conclusions}

We have shown that the equivalence between the two-dimensional Dirac oscillator and the anti-Jaynes-Cummings model extends to a q-deformed 
scenario. After this fundamental equivalence we have investigated from first principles the Hamiltonian spectrum and time evolution for initial number
and coherent states, focusing on the appearance of the {\it Zitterbewegung} effect. We have highlined the algebraic  modifications introduced by the 
q-deformation. In particular we have shown the lack of preservation of the q-deformed versions of the total angular momentum.  Actually we have 
shown it leads to a time evolution mimicking the well-known collapse-revival structure of the population inversion or spin in standard undeformed 
Jaynes-Cummings model. Moreover, we have provided suitable relations for the non relativistic limit. 

\section*{Acknowledgments}

A. L. acknowledges financial support from Spanish Ministerio de Econom\'ia y Competitividad 
Project No. FIS2016-75199-P, and from the Comunidad Aut\'onoma de Madrid research  consortium QUITEMAD+ 
Grant No. S2013/ICE-2801.

\appendix

\smallskip

\section{Differential forms}

Here we provide explicit differential expressions  for the Hamiltonians of the q-deformed  Dirac oscillator and the Jaynes-Cummings and anti-Jaynes Cummings models.
For the Dirac oscillator, we have after Eqs. (\ref{df}) and (\ref{HDq}):  

\begin{widetext}

\begin{multline}\label{10}
 H^D_{q}={mc^2\sigma_{z}}+G[\sigma_{x}[(2\cos(2x\alpha)+2 \sin(2y\alpha)-\cos(x\alpha+\alpha{d\over dx})-i \sin(x\alpha+\alpha{d\over dx})  \\
+{i\cos(y\alpha+\alpha{d\over dy})- \sin(y\alpha+\alpha{d\over dy})-\cos(\alpha {d\over dx}-x\alpha)
}-i \sin(\alpha {d\over dx}-x\alpha)\\
-i\cos(\alpha {d\over dy}-y\alpha)+ \sin(\alpha {d\over dy}-y\alpha)]+\sigma_{y}[-2 \sin(2x\alpha)+2\cos(2y\alpha)
-i\cos(x\alpha+\alpha{d\over dx})\\
-{\cos(y\alpha+\alpha{d\over dy})+ \sin(x\alpha+\alpha{d\over dx})-i \sin(y\alpha+\alpha{d\over dy})}
+i\cos(\alpha {d\over dx}-x\alpha)- \sin(\alpha {d\over dx}-x\alpha)\\
-\cos(\alpha {d\over dy}-y\alpha)-i \sin(\alpha {d\over dy}-y\alpha)]],
\end{multline}
where G satisfies the relation
\begin{equation}\label{11}
G=c{{{\sqrt{\hbar m\omega}}}\over {\sqrt{2(1-e^{-2{\alpha^2}} )}}}   .
\end{equation}

For the JC Hamiltonian (\ref{HJCq}):
\begin{eqnarray}\label{15}
H_{q,i}^{JC} &=&\frac{\hbar \eta^\prime}{ 2 \sqrt{1-e^{-2 \alpha^2 }}}{[\sigma_{x}[-2 \sin(\phi^\prime -2r_{i}\alpha)-i\cos(\phi^\prime-r_{i}\alpha+\alpha{d\over dr_{i}})+ \sin(\phi^\prime-r_{i}\alpha+\alpha{d\over dr_{i}})} \nonumber \\
& &+{i\cos(\phi^\prime-\alpha {d\over dr_{i}}-r_{i}\alpha)+ \sin(\phi^\prime-\alpha {d\over dr_{i}}-r_{i}\alpha)]
+}\left([\sigma_{y}[-2\cos(\phi^\prime-2\alpha r_{i}\right)
\nonumber\\
& &{-i \sin(\phi^\prime-\alpha {d\over dr_{i}}-r_{i}\alpha)+i \sin(\phi^\prime+\alpha {d\over dr_{i}}-r_{i}\alpha)}+\cos(\phi^\prime-\alpha {d\over dr_{i}}-r_{i}\alpha)
\nonumber\\
& &+{\cos(\phi^\prime+\alpha {d\over dr_{i}}-r_{i}\alpha }\left )])] + {(\hbar\delta^\prime \sigma_{z}}\right) ,
\end{eqnarray}
where $r_{i=x,y} = x,y$, while for the A JC Hamiltonian (\ref{HAJCq}):
\begin{multline}\label{17}
 H_{q,i}^{AJC}=\frac{\hbar \eta}{ 2 \sqrt{1-e^{-2 \alpha^2}}} [\sigma_{x}(2 \sin(\phi+ 2r_{i}\alpha)+i\cos(\phi+r_{i}\alpha+\alpha{d\over dr_{i}})- \sin(\phi+r_{i}\alpha+\alpha{d\over dr_{i}})\\
-{i\cos(\phi+\alpha {d\over dr_{i}}+r_{i}\alpha)- \sin(\phi-\alpha {d\over dr_{i}}+r_{i}\alpha)]
+}[\sigma_{y}(2\cos(\phi+2\alpha r_{i}){-i \sin(\phi+\alpha {d\over dr_{i}}+r_{i}\alpha)}\\
-\cos(\phi+\alpha {d\over dr_{i}}+r_{i}\alpha)
-{\cos(\phi-\alpha {d\over dr_{i}}+r_{i}\alpha)+
+}{{i \sin(\phi-\alpha {d\over dr_{i}}+r_{i}\alpha}]]} +  \hbar\delta  \sigma_{z} .
\end{multline}

\end{widetext}


\begin{thebibliography}{11}

\bibitem{IMC67}
D. It\^{o}, K. Mori and E. Carrieri, 
An example of dynamical systems with linear trajectory,
Nuovo Cimento {\bf 51A}, 1119 (1967).

\bibitem{MS89}
M. Moshinsky and A. Szczepaniak, 
The Dirac oscillator,
J. Phys. A 22, L817 (1989).

\bibitem{BMS07}
A. Bermudez, M. A. Martin-Delgado, and E. Solano, 
Exact mapping of the 2+1 Dirac oscillator onto the Jaynes-Cummings model: Ion-trap experimental proposal,
Phys. Rev. A {\bf 76}, 041801(R) (2007).

\bibitem{BML08}
A. Bermudez, M.A Martin-Delgado, and A. Luis, 
Nonrelativistic limit in the 2+1 Dirac oscillator: A Ramsey-interferometry effect,
Phys. Rev. A {\bf 77},  033832 (2008).

\bibitem{FSBKMS13}
J. A. Franco-Villafa\~{n}e, E. Sadurn\'{\i}, S. Barkhofen, U. Kuhl, F. Mortessagne, and T. H. Seligman,
First Experimental Realization of the Dirac Oscillator,
Phys. Rev. Lett. {\bf 111}, 170405 (2013).

\bibitem{SL10}
S. Longhi,
Photonic realization of the relativistic Dirac oscillator,
Opt. Lett. {\bf 35}, 1302 (2010).

\bibitem{LCLJ11}
Y. Luo, Y. Cui, Z. Long, and J. Jing,
2+1 Dimensional Noncommutative Dirac Oscillator and (Anti)-Jaynes-Cummings Models,
Int. J. Theor. Phys. {\bf 50},  2992  (2011).

\bibitem{JMD12}
A. Jellal, A. El Mouhafid and M. Daoud,
Massless Dirac fermions in an electromagnetic field,
J. Stat. Mech. {\bf 2012},  P01021 (2012). 

\bibitem{JCM}
E. T. Jaynes and W. F. Cummings,
Comparison of quantum and semiclassical radiation theories with application to the beam maser,
Proc. IEEE {\bf 51}, 89  (1963);
B. W. Shore and P. L. Knight ,
The Jaynes-Cummings Model,
J. Mod. Opt. {\bf 40}, 1195 (1993).

\bibitem{LBMW03}
D. Leibfried, R. Blatt, C. Monroe, and D. Wineland,
Quantum dynamics of single trapped ions,
Rev. Mod. Phys. {\bf 75}, 281(2003).

\bibitem{qD} 
A. J. Macfarlane, 
On q-analogues of the quantum harmonic oscillator and the quantum group $SU(2)_q$,
J. Phys. A {\bf 22}, 4581 (1989);
L. S. Biedenharn, 
The quantum group $SU_q (2)$ and a q-analogue of the boson operators,
J. Phys. A {\bf  22}, L873 (1989);
P. P. Kulish and E. V. Damaskinsky, 
On the q oscillator and the quantum algebra $su_q (1,1)$,
J. Phys. A {\bf 23}, L415 (1990).

\bibitem{LRW97}
A. Lorek, A. Ruffing and J. Wess, 
A q-Deformation of the Harmonic Oscillator,
Z. Phys. C {\bf 74}, 369 (1997).

\bibitem{EM06}
V.V. Eremin, A.A. Meldianov, 
The q-deformed harmonic oscillator, coherent states, and the uncertainty relation,
Theor. Math. Phys. {\bf 147},  709  (2006).

\bibitem{HS16}
N. Hatami and M. R. Setare,
The q-deformed Dirac oscillator in 2 + 1 dimensions,
Phys. Lett. A {\bf 380},  3469  (2016).

\bibitem{VD02}
V. V. Dodonov, 
`Nonclassical' states in quantum optics: a `squeezed' review of the first 75 years,
J. Opt. B: Quantum Semiclass. Opt. {\bf 4}, R1 (2002).

\bibitem{GR00}
W. Greiner, 
{\it Relativistic Quantum Mechanics: Wave Equations} (Springer, Berlin, 2000).

\bibitem{FI91}
D. I. Fivel,
Quasi-coherent states and the spectral resolution of the q-Bose field operator, 
J. Phys. A {\bf 24}, 3575  (1991);
V. Spiridonov, Coherent states of the q-Weyl algebra, Lett. Math. Phys. {\bf 35}, 179 (1995). 

\bibitem{CDG04}
C. Cohen-Tannoudji, J. Dupont-Roc, and G. Grynberg,
{\it Atom-Photon Interactions. Basic Processes and Applications}, (Wiley-VCH, Weinheim, 2004).

\bibitem{CMM94}
E. A. Kochetov,
Exactly solvable non-linear generalisations of the Jaynes-Cummings model, 
J. Phys. A: Math. Gen. {\bf 20},  2433 (1987);
J. \v{C}rnugelj, M. Martinis, and V. Mikuta-Martinis, 
Properties of a deformed Jaynes-Cummings model,
Phys. Rev. A {\bf 50}, 1785 (1994).

\bibitem{CR11}
S. Cordero and J. R\'{e}camier,
Selective transition and complete revivals of a single two-level atom in the JaynesÐCummings Hamiltonian with an additional Kerr medium,
J. Phys. B: At. Mol. Opt. Phys. {\bf 44}, 135502  (2011) ;
O. de los Santos-S\'{a}nchez and J. R\'{e}camier, 
The f-deformed JaynesÐCummings model and its nonlinear coherent states,
J. Phys. B: At. Mol. Opt. Phys. {\bf 45}, 015502 (2012).

\bibitem{VMF95}
W. Vogel and R. L. de Matos Filho, 
Nonlinear Jaynes-Cummings dynamics of a trapped ion,
Phys. Rev. A {\bf 52}, 4214 (1995); 
R. L. de Matos Filho and W. Vogel, 
Nonlinear coherent states,
Phys. Rev. A {\bf 54}, 4560 (1996).

\bibitem{RL95}
R. Lynch, 
The quantum phase problem: a critical review, 
Phys. Rep. {\bf 256}, 367 (1995);
V. Pe\v{r}inov\'{a},  A. Luk\v{s} and J. Pe\v{r}ina, 
{\it Phase in Optics}  (World Scientific, Singapore, 1998);
P. Carruthers and M. M. Nieto, 
Phase and angle variables in quantum mechanics, 
Rev. Mod. Phys. {\bf 40}, 411 (1968);
D. T. Pegg and  S. M. Barnett, 
Quantum optical phase, 
J. Mod. Opt. \textbf{44}, 225 (1997).

\end{thebibliography}
\end{document}